\documentclass[preprint]{iucr}              
\usepackage{xcolor}
\usepackage{xspace}
\usepackage{graphicx}
\usepackage{epstopdf, epsfig}
\usepackage{mathtools, amsmath, amssymb, amsfonts}
\usepackage{algorithm, algpseudocode}

\definecolor{dgreen}{HTML}{008000}

\newcommand{\fig}[1]{Fig.~\ref{fig:#1}}
\newcommand{\tabl}[1]{Table~\ref{tab:#1}}
\newcommand{\etal}{\textit{et al.}\xspace}
\newcommand{\qmin}{$Q_\text{min}$\xspace}
\newcommand{\qmax}{$Q_\text{max}$\xspace}
\newcommand{\qdamp}{$Q_\text{damp}$\xspace}
\newcommand{\qbroad}{$Q_\text{broad}$\xspace}
\newcommand{\rmax}{$r_\text{max}$\xspace}
\newcommand{\rmin}{$r_\text{min}$\xspace}
\newcommand{\rgrid}{$r_\text{grid}$\xspace}
\newcommand{\uiso}{$U_\text{iso}$\xspace}
\newcommand{\sgmining}{{\sc spacegroupMining} \xspace}

\graphicspath{{./plot/}}

     \journalcode{J}              

\begin{document}



\title{Robustness test of the \textbf{\textit{spacegroupMining}} model for determining space groups from atomic pair distribution function data}
\keyword{Robustness Test, Machine Learning, Data Mining, Space Group, Pair Distribution Function}

\author[a]{Ling}{Lan}{}{}
\cauthor[a]{Chia-Hao}{Liu}{}{}
\cauthor[a, b]{Qiang}{Du}{}{}
\cauthor[a, c]{Simon~J.~L.}{Billinge}{sb2896@columbia.edu}{}
\aff[a]{Department of Applied Physics and Applied Mathematics, Columbia University, \city{New York}, NY 10027, \country{United States}}
\aff[b]{Data Science Institute, Columbia University, \city{New York}, NY 10027, \country{United States}}
\aff[c]{Condensed Matter Physics and Materials Science Department, Brookhaven National Laboratory, Uptown, NY 11973, \country{United States}}

\shortauthor{Lan \etal}

\maketitle                        

\begin{abstract}
Machine learning models based on convolutional neural networks have been used for predicting space groups of crystal structures from their atomic pair distribution function (PDF). However, the PDFs used to train the model are calculated using a fixed set of parameters that reflect specific experimental conditions, and the accuracy of the model when given PDFs generated with different choices of these parameters is unknown. In this paper, we report that the results of the top-1 accuracy and top-6 accuracy are robust when applied to PDFs of different choices of experimental parameters \rmax, \qmax, \qdamp and atomic displacement parameters.
\end{abstract}


\section{Introduction}

Recently it was shown \cite{liu;aca19} that a convolutional neural network (CNN) machine learning model could predict the space group of a material from its atomic pair distribution function (PDF) \cite{billi;b;itoch19,egami;b;utbp12} with good accuracy.  This model is called \sgmining and was recently deployed as a web application on the pdfitc.org website \cite{yang;aca21}.

The atomic pair distribution function (PDF) method is a total scattering technique for determining local order in nanostructured materials. Theoretically, the PDF gives the scaled probability of finding two atoms in a material a distance $r$ apart and is related to the density of atom pairs in the material \cite{billi;b;itoch19,egami;b;utbp12}.

\begin{figure}
\includegraphics[width=0.75\columnwidth]{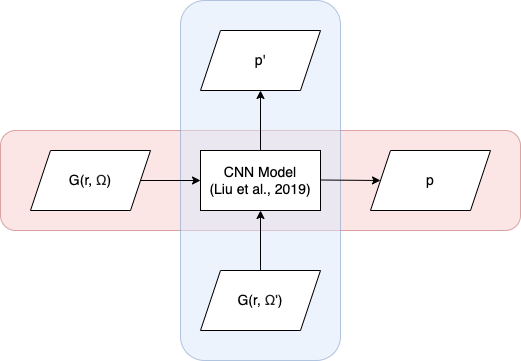}
\caption{$G(r,\Omega)$ are the calculated PDFs w.r.t. the original experimental parameters used to train the CNN model, while $G(r,\Omega')$ are PDFs with varying parameters. $p$ and $p'$ are the corresponding model outputs from which we could measure the accuracy of the model predictions. The red (blue) section depicts the training (testing) process respectively.}
\label{fig:flowchart}
\end{figure}

The model of \cite{liu;aca19} was trained, as shown in the red section in Figure \ref{fig:flowchart}, using calculated PDFs, denoted by $G(r,\Omega)$ here, where $\Omega$ indicates the set of parameters that define experimental details of the measurement and the sample.  These model experimental parameters that affect the quality of the data, such as the maximum range of $Q$, \qmax, where $Q$ is the modulus of the scattering vector \cite{egami;b;utbp12}, the $r$-range of the input PDF, \rmax, and so on \cite{proff;jac99}. These are listed in full in \tabl{param}. Although a specific set of values were used to train the model, in general, different parameter values might be relevant in a scientist's measured PDFs. We denote these as $G(r,\Omega')$, where the prime on the $\Omega$ indicates that some experimental values in the set are different from the ones used in the training. A natural and important question is whether the trained model with $G(r,\Omega)$ could provide reasonable predictions on materials associated with $G(r,\Omega')$. If the accuracy of the model predictions on $G(r,\Omega')$ is close to its performance on the PDFs, $G(r,\Omega)$, that the model learned from, we believe that the model is robust. To be more explicit, this robustness test investigates how input data quality translates to performance, while the input data distribution does not shift, i.e., the materials we use to train and test are not varying. In this paper we assess how well the model performs when it is tested on PDFs that were calculated using experimental parameters different from those for the training set (blue section in Figure \ref{fig:flowchart}). We conclude that overall it performs quite well with respect to \rmax, \qmax, \qdamp and atomic displacement factor (ADP), or \uiso, of the measurement, hence providing evidence to the robustness of the CNN machine learning model developed in  \cite{liu;aca19}.

\section{Method}

\subsection{Data and Model \label{sec:Model}}

Our main objective is to test the robustness of the originally trained model. However, we are not able to identify the exact datasets that constituted the training set in the original training. In order to avoid testing robustness by inadvertently using a dataset that might be part of the original model's training data, our first step is to rebuild the model again.

The input PDF data are calculated from $98,830$ structures in the~45 most heavily represented space groups in the ICSD \cite{belsky;acb02} structural database. The PDFs are calculated from crystallographic information framework (CIF) \cite{hall;aca91} files obtained from ICSD using the diffpy-cmi \cite{juhas;aca15} package with parameters $\Omega$ defined in Table~\ref{tab:param}. The parameters are the same as the ones used in \cite{liu;aca19}, except that the grid size is $\pi/40$ in our experiment (the paper used $\pi/23$), so that we could calculate the PDF with higher \qmax's. $80\%$ of the data is considered as training samples, and the rest is treated as test samples. The choice of \rmin, \rmax, and \rgrid in Table~\ref{tab:param} discretizes the input PDFs to 1D signal sequences of dimension $209\times1\times1$. We further normalize the PDF input, $G(r,\Omega)$, to ensure that it lies between 0 and 1 for each entry.
\begin{table}
\begin{center}
\caption{Experimental parameters used to calculate the PDFs used for training, validation and testing of the model. Here \qmin and \qmax define the range of $Q$ that was included in the Fourier transform to obtain the PDF ($Q$ is the modulus of the scattering vector), \rmin and \rmax are the minimum and maximum range of $r$ of the PDF and \rgrid was the size of the bins in the numerical determination of $G$. \uiso is the atomic displacement factor of the atoms.  \qdamp and \qbroad are parameters that determine the $r$-dependence of the damping of the PDF signal, mostly due to the $Q$-space resolution of the measurement.  All these parameters are defined in detail in \cite{egami;b;utbp12,proff;jac99}.}
\begin{tabular}{|c|c|}
\hline
\rmin (\AA) & 1.5\\\hline
\rmax (\AA) & 30.0\\\hline
\rgrid (\AA) & $\pi/40$\\\hline
\qmin (\AA$^{-1}$) & 0.5\\\hline
\qmax (\AA$^{-1}$) & 23.0\\\hline
\uiso (\AA$^{2}$) & 0.008\\\hline
\qdamp (\AA$^{-1}$) & 0.04\\\hline
\qbroad (\AA$^{-1}$) & 0.01\\\hline
\end{tabular}
\end{center}
\label{tab:param}
\end{table}

To rebuild the model, we use the architecture based on the convolutional neural network (CNN) used in Liu's paper \cite{liu;aca19}. The output, $p$, of the model is a $45\times1$ vector, which represents the probability of the input PDF being in each of the $45$ space groups considered in our study. We use weighted categorical cross entropy loss,
\begin{align}\label{eq:loss}
\text{Loss} = -\sum_{i=1}^{45} w_i\cdot {p_{true}}_i\cdot \log p_i,
\end{align}
to mitigate the effects of unbalanced data, where the weight $w_i$ is defined as the number of structures in the training set over the number of structures of each space group in the training dataset. Adaptive moment estimation (Adam) with a mini-batch size of $64$ is used to train the model. Furthermore, we modify the learning rate as an exponential decay, $l=5\times10^{-4}e^{-0.025\times\text{epoch}}$. The model is trained using Keras on a single Nvidia Tesla P100 GPU.

An accuracy of $67.7\%$ from top-1 prediction and $90.2\%$ from top-6 predictions is achieved. The performance of our reconstructed model is similar to the one shown in the original paper, which was $70.0\%$ top-1 accuracy and $91.9\%$ top-6 accuracy. The model rebuilt here is used, without any further retraining, in subsequent robustness tests on datasets involving PDFs having different parameter values, as illustrated schematically in \fig{flowchart}.

\subsection{Robustness Test}

In order to test the robustness, we consider four experimental parameters that are used to calculate the PDFs from the structural CIFs, which are \qmax, \rmax, \qdamp, and \uiso. The other parameters in \tabl{param} are not expected to affect the accuracy greatly and were not explicitly tested. Variations in \qmin produce no effect until low-$Q$ Bragg peaks are lost and then result in long-wavelength damped sinusoidal oscillations in the PDF that appear like an oscillating background to the signal. The data are interpolated onto a different \rgrid during the process and so the user \rgrid will not affect the outcome, and most users are expected to have data calculated to an arbitrarily small \rmin.  Finally, \qbroad has a very small effect on the width of peaks in rather high $r$-regions that are unlikely to be uploaded.  

To carry out the tests we randomly choose structures from the testing set ($10\%$ of the testing samples are chosen), and compute their PDFs while varying each of these parameters separately between limits that are chosen to bracket values that are experimentally reasonable. These calculated PDFs are then given to the trained model, without model retraining despite the changes in parameter values of the input PDFs, to predict the space group, and the model accuracy is computed as a function of the experimental parameter value.

First we consider the robustness against a variation in \rmax. The model was trained with an \rmax of 30~\AA\ and we want to test its performance when given PDFs computed (or measured) over a narrower $r$-range, from 10~\AA\ to 30~\AA\ every 2~\AA.  Variations in \rmax will change the length of the PDF vector, which is not allowed in our model. Since we are only considering $r$-ranges that are shorter than 30~\AA, to keep the dimension of all input PDFs consistent, the data are padded with zero's up to the value of \rmax $= 30$~\AA\ before being interpolated on to the $209\times1\times1$ grid using quadratic interpolation.

To test the \qmax sensitivity, computed PDFs in the range of $12 \leq$ \qmax $\leq 30$~\AA$^{-1}$ in steps of 3~\AA$^{-1}$ were tested against the trained model. For \qdamp, we tested on computed PDFs in the range of $0 \leq$ \qdamp $\leq 0.15$~\AA$^{-1}$ in steps of $0.03$~\AA$^{-1}$. Finally, for the ADP, \uiso, from $0.005 \leq$ \uiso $\leq 0.01$~\AA$^{2}$ in steps of 0.001~\AA$^{2}$, where the model was trained on values \qmax $= 23$~\AA$^{-1}$, \qdamp $= 0.04$~\AA$^{-1}$ and \uiso $= 0.008$~\AA$^{2}$, respectively.

\section{Results \label{sec:Robustness-Test-Results}}

\subsection{Robustness with respect to \rmax}

\begin{figure}
\includegraphics[width=0.75\columnwidth]{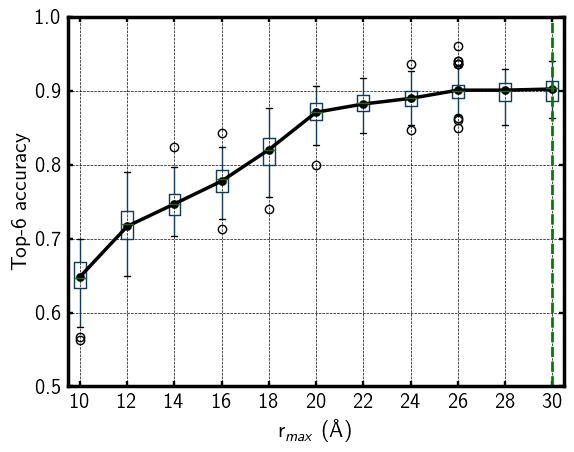}
\caption{The black dots represent the top-6 accuracy as \rmax is varied on the testing datasets. The value of \rmax $= 30$~\AA\ used to train the model is shown as a vertical green dashed line. The box plot at each \rmax value shows the uncertainties on the top-6 accuracy (see text for details). The box extends from the first quartile (Q1) to the third quartile (Q3) of the distribution, with a green horizontal line at the median. The whiskers from the box extend by 1.5 times the inter-quartile range (IQR). The hollow circles indicate samples that lie outside that range.}
\label{fig:rmax}
\end{figure}

Figure \ref{fig:rmax} shows the top-6 accuracy against a variation in \rmax from 10~\AA\ to 30~\AA. When \rmax is larger than 20~\AA, top-6 accuracy is always above $87.1\%$, which is close to the optimal value of $90.2\%$. It is recommended to give the model a PDF with a $r_{max}\geq 30$~\AA. However, the robustness test shows that if the signal is from data over a narrower range, such as a nanoparticle whose signal dies on a shorter length-scale, the model can still be categorized into space group with reasonably good accuracy, though the performance drops off more quickly below an \rmax of 20~\AA\ or so.

\subsection{Robustness with respect to \qmax}

\begin{figure}
\centering
\includegraphics[width=0.75\columnwidth]{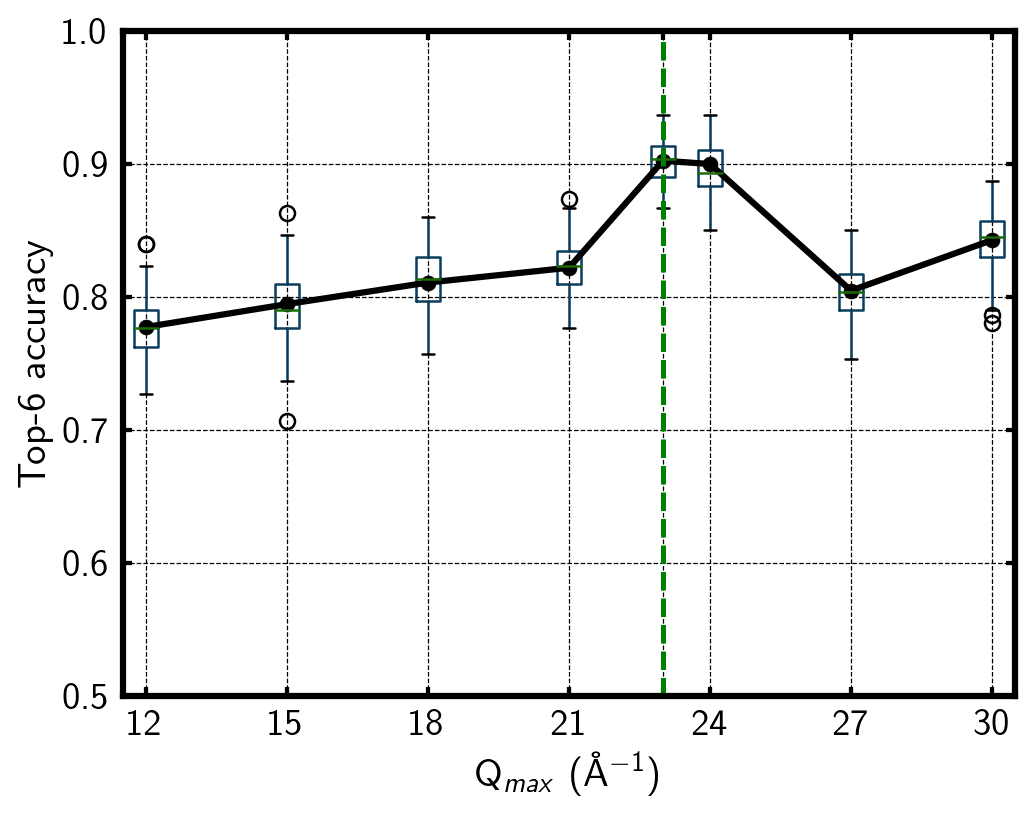}
\caption{The black dots represent the top-6 accuracy as \qmax is varied. The value of \qmax $= 23$~\AA$^{-1}$ used to train the model is shown as a vertical green dashed line. The box plot at each \qmax value is plotted in the same way as the ones in Fig.~\ref{fig:rmax}.}
\label{fig:qmax}
\end{figure}

Next we consider the robustness of the model when PDFs are generated using different \qmax values. As shown in Figure \ref{fig:qmax}, when \qmax is larger than 18~\AA$^{-1}$, top-6 accuracy is above $81.1\%$. The bump around 23 \AA$^{-1}$ makes sense, as the model favors the \qmax value that it is trained on. But the performance with \qmax values deviated from 23~\AA$^{-1}$ is still fairly good over the entire range of values considered, the accuracy never falls below $77.7\%$, and so the model is quite robust against variations in \qmax.

\subsection{Robustness with respect to \qdamp}

\begin{figure}
\centering
\includegraphics[width=0.75\columnwidth]{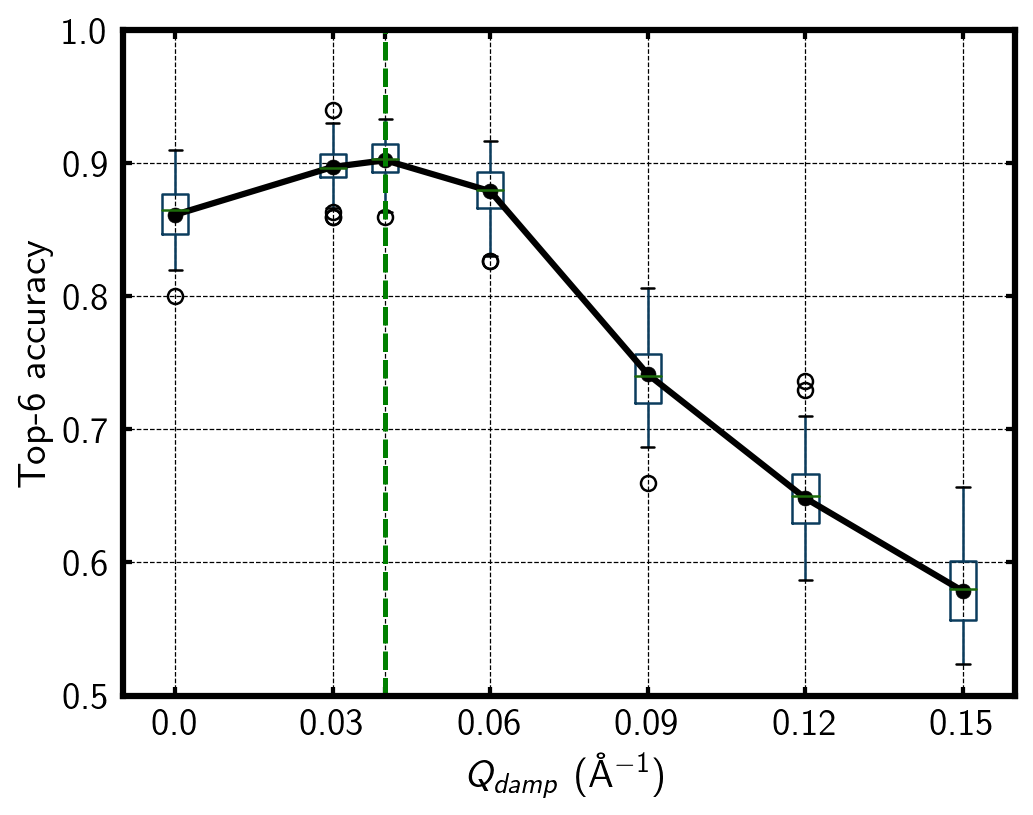}
\caption{The black dots represent the top-6 accuracy as \qdamp is varied. The value of \qdamp $= 0.04$~\AA$^{-1}$ used to train the model is shown as a vertical green dashed line. The box plot at each \qdamp value is plotted in the same way as the ones in Fig.~\ref{fig:rmax}.}
\label{fig:qdamp}
\end{figure}

Figure~\ref{fig:qdamp} shows the top-6 accuracy against a variation in \qdamp in the range of $0 \leq Q_\text{damp} \leq 0.15$~\AA$^{-1}$ in steps of $0.03$~\AA$^{-1}$.
When \qdamp is smaller than $0.06$~\AA$^{-1}$, the top-6 accuracy is always above $86.1\%$. However, the performance drops off fairly quickly above \qdamp$=0.06$~\AA$^{-1}$ or so. When \qdamp$=0.15$~\AA$^{-1}$, the PDF signal practically vanishes in the region above $r=20$~\AA\ and so we might expect the accuracy to be similar to that of \rmax$=20$~\AA. We find that the accuracy of \qdamp$=0.15$~\AA$^{-1}$ falls to $57.8\%$, which is significantly lower than the value of $87.1\%$ of \rmax$=20$~\AA.  This is presumably because \qdamp damps the signal progressively over the entire range and therefore the model is more sensitive to \qdamp variations than \rmax. However, we note that the accuracy with \qdamp values deviated from $0.06$~\AA$^{-1}$ never falls below $57.8\%$, which can still give acceptable results in many cases.

\subsection{Robustness with respect to Atomic Displacement Parameter, \uiso}

\begin{figure}
\centering
\includegraphics[width=0.8\columnwidth]{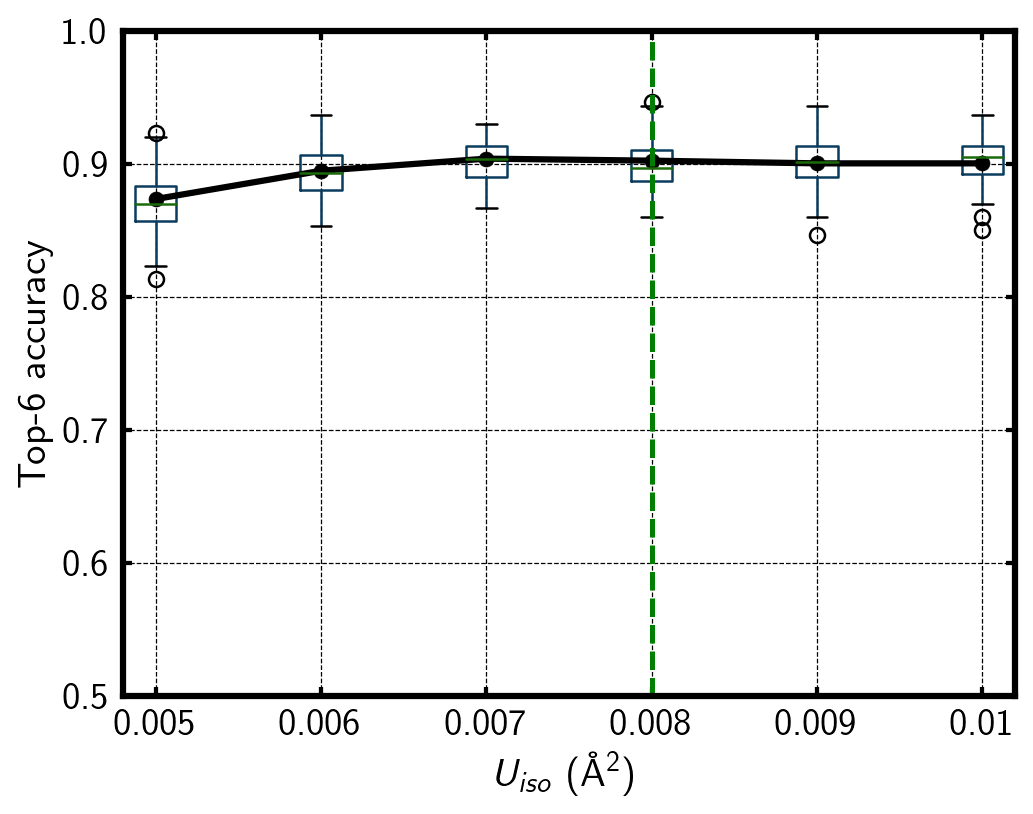}
\caption{The black dots represent the top-6 accuracy as \uiso is varied. The value of \uiso $= 0.008$~\AA$^{2}$ used to train the model is shown as a vertical green dashed line. The box plot at each \uiso value is plotted in the same way as the ones in Fig.~\ref{fig:rmax}.}
\label{fig:adp}
\end{figure}

Finally (Fig.~\ref{fig:adp}), we consider robustness against variations in \uiso. The results are even less sensitive to the choice of ADP. When \uiso of the PDFs were in the range 0.005~\AA$^{2}$ to 0.01~\AA$^{2}$, the top-6 accuracy is always above $87.3\%$.

The numbers from all the robustness tests are reproduced in the supplementary information associated with this paper.

\section{Discussion}

The main goal of this work is to investigate the model performance when given PDFs computed using different experimental parameters than the model was trained with.  We found that the model predictions are quite robust for reasonable ranges of parameters, giving hope that the tool can be used by experimentalists when they have PDFs measured under different conditions.  Here we speculate on some more detailed aspects of the findings.

Two conditions must be satisfied for the model to perform well.  First, the input data must contain sufficient information to do the differentiation by space group.  Second, the values of the learned CNN model parameters must be able to correctly classify based on this information, even when it has been distorted by the use of different experimental parameters.   We briefly discuss the effect of each experimental parameter on the information content of the PDF. These parameters are discussed in detail in \cite{egami;b;utbp12}. Lowering \qmax can result in a loss of information from the missing high-$Q$ region, and results in a broadening of the peaks in the PDF.  Increases in \uiso also broaden the PDF peaks but are coming from increased static or thermal disorder in the sample itself. Because of peak overlap in the PDF, especially in the high-$r$ region, broadened peaks always result in a loss of structural information. Clearly, lowering the range of data used (\rmin to \rmax) decreases the information content of the supplied data.  These are parameters under the control of the experimenter though, as they are parameters that they can set during the data analysis to produce the PDF.  A  lower $Q$-space resolution of the measurement (resulting in a higher \qdamp) can also lower the information content of the data due to Bragg peak overlap, especially in the high-$Q$ region of the data before it is Fourier transformed to obtain the PDF. In the PDF this appears as an approximately Gaussian fall off in the structural signal with increasing-$r$.  Any intrinsic nanocrystallinity in the sample, such as finite nanoparticles or loss of structural coherence in the form of smaller crystallites or domains, has a similar effect as \qdamp on the PDF.  The fall-off in the signal will have a different functional form in this case (for example a power-law in the case of spherical domains/particles) but roughly speaking appears in a rather similar way as the Gaussian dropoff modelled by \qdamp and so we did not explicitly separate these factors in this robustness test.  We tested \qdamp over a range of values that simulated structural coherence down to a $\sim 2$~nm diameter. Finally, \qmin and \qbroad are expected to have only a very small effect on the accuracy.  \qbroad is only relevant for data with very asymmetric Bragg peaks, for example, coming from time-of-flight neutron data, and even then, only at very high values of~$r$ that tend to be higher than the values we have been giving to the CNN.  \qmin is often determined by the shadow of the beamstop in an experiment and will only affect the data if any low-angle signal is lost due to this.  In that case, it results in very long wavelength undulations in the background of the resulting PDF that will not affect the model's ability to classify by spacegroup.

The observed robustness indicates that measured PDFs generally contain sufficient information to make this space-group determination, even when the data content is degraded somewhat by reduced real-space resolution (lower \qmax higher \uiso) and a more limited $r$-range of the data (lower \rmax and higher \qdamp).  The accuracy falls off more rapidly when there is a loss of information in the PDF (lower \qmax, \rmax, higher \qdamp, \uiso); however, we note that the accuracy also falls off when we give the model a dataset with higher resolution or $r$-range, and therefore increased information content.  The fall-off in accuracy in these cases must be due to the less than optimal learned CNN parameter values.  This could be addressed by retraining the model with a wider range of experimental parameters, but it seems that it may not be required, except perhaps for the case of small nanoparticles (represented by large \qdamp values above~0.08 in this study).  \rmin and \rmax is largely under the control of the experimentalist (it is a setting in most PDF data analysis programs), but more importantly, the range of $r$-that the signal persists over depends on the crystallite/domain size of the sample and whether it is nanocrystalline.  This suggests that training a new model suitable for small nanoparticles (i.e., data signal ranges up to 1 or 2~nm) may be warranted.  We will look into this in the future and consider deploying it at the PDFitc website.

Another way that information in data is degraded is the presence of noise.  Noise may be random or correlated.  We have not systematically tested the robustness of the model to the presence of added noise in the data because it is difficult to reliably mimic the actual errors that are present in real data.  A more meaningful measure of this is to establish how well the model works on actual datasets from known materials.  This was reported in the original paper \cite{liu;aca19}. There it was found that of 15 experimental PDFs the model gave a correct prediction in the top-6 from 12 cases.  This is not a large sample, but is an 80\% accuracy.  Given that the datasets were already obtained with experimental parameters that are not necessarily those that the model was trained with, this is comparable, if somewhat degraded, performance to the test data without noise that we report here.  The \sgmining model is apparently also quite robust against the effects of measurement noise.

\section{Conclusions}

The use of deep learning to do complex classifications from data is a potentially useful approach that is becoming more widespread in materials science, crystallography and diffraction.  Inherent in the process is that the model was trained on a particular set of data and its applicability to do the classification on data that is, in some way, different, for example, measured with different resolutions or over different ranges, might limit its ability to make accurate predictions.  In general, the model may be retrained on a wider set of data that incorporates cases of different ranges, resolutions and so on.  However, here, for the case of the \sgmining model that is deployed on pdfitc.org, we simply explored its robustness in making accurate predictions on different range and resolution data without retraining the model.  The main result is that the model is quite robust and performs well without having to be retrained in most cases.  Modest reductions in prediction accuracy were observed, but it still performed well given a rather wide, but reasonable, range of resolution and range parameters, suggesting that it is not of great urgency to retrain it.  We note that retraining it with a more diverse set of training data, whilst increasing accuracy for parameter values away from the original training values, it may decrease the prediction accuracy for PDFs with the original set of parameter values, where those values were chosen as being somewhat representative of values in many rapid acquisition x-ray PDF studies. Through this work, it has been shown that, without additional retraining, the spacegroupMining@pdfitc model still performs with reasonable accuracy for a relatively wide range of experimental parameters, and can thus be used as a robust computational tool.


\ack{Funding information}

This work in the Billinge group was supported by the U.S. National Science Foundation through grant DMREF-1922234 and CCF-1704833.

\pagebreak
\begin{center}
\textbf{\large Supplemental Materials}
\end{center}

\begin{table}
\centering
\caption{Top-6 accuracy and top-1 accuracy when \rmax is chosen from 10~\AA\ to 30~\AA.}
\label{table:rmax}
\begin{center}
\begin{tabular}{|c|c|c|c|c|c|c|c|c|c|c|c|}
\hline
\rmax (\AA) & 10 & 12 & 14 & 16 & 18 & 20 & 22 & 24 & 26 & 28 & \textbf{30}\\\hline
\hline
Top-6 accuracy &0.648	&0.717	&0.747	&0.778	&0.820	&0.871	&0.882	&0.890	&0.901	&0.901	&\textbf{0.902}\\\hline
Top-1 accuracy &0.285	&0.367	&0.433	&0.449	&0.511	&0.552	&0.600	&0.617	&0.652	&0.671	&\textbf{0.677}\\\hline
\end{tabular}
\end{center}
\end{table}

\begin{table}
\centering
\caption{Top-6 accuracy and top-1 accuracy when \qmax is chosen from 12~\AA$^{-1}$ to 30~\AA$^{-1}$.}
\label{table:qmax}
\begin{center}
\begin{tabular}{|c|c|c|c|c|c|c|c|c|}
\hline
\qmax (\AA$^{-1}$) & 12 & 15 & 18 & 21 & \textbf{23} & 24 & 27 & 30\\\hline
\hline
Top-6 accuracy
&0.777	&0.795	&0.811	&0.822	&\textbf{0.902}	&0.900	&0.805	&0.84\\\hline
Top-1 accuracy
&0.516	&0.591	&0.597	&0.604	&\textbf{0.677}	&0.663	&0.598	&0.610\\\hline
\end{tabular}
\end{center}
\end{table}

\begin{table}
\centering
\caption{Top-6 accuracy and top-1 accuracy when \qdamp is chosen from 0~\AA$^{-1}$ to $0.15$~\AA$^{-1}$.}
\label{table:qdamp}
\begin{center}
\begin{tabular}{|c|c|c|c|c|c|c|c|}
\hline
\qdamp (\AA$^{-1}$) & 0 & 0.03 & \textbf{0.04} & 0.06 & 0.09 & 0.12 & 0.15\\\hline
\hline
Top-6 accuracy
&0.861 &0.897 &\textbf{0.902} &0.879 &0.741 &0.648 &0.578\\\hline
Top-1 accuracy
&0.602 &0.659 &\textbf{0.677} &0.579 &0.390 &0.294 &0.234\\\hline
\end{tabular}
\end{center}
\end{table}

\begin{table}
\centering
\caption{Top-6 accuracy and top-1 accuracy when ADP is chosen from 0.005~\AA$^{2}$ to 0.01~\AA$^{2}$.}
\label{table:adp}
\begin{center}
\begin{tabular}{|c|c|c|c|c|c|c|}
\hline
adp (\AA$^{2}$) & 0.005 & 0.006 & 0.007 & \textbf{0.008} & 0.009 & 0.01\\\hline
\hline
Top-6 accuracy
&0.873	&0.895	&0.904	&\textbf{0.902}	&0.900	&0.900\\\hline
Top-1 accuracy
&0.618	&0.649	&0.664	&\textbf{0.677}	&0.666	&0.641\\\hline
\end{tabular}
\end{center}
\end{table}

\end{document}